\documentclass[
reprint,
superscriptaddress,
twocolumn,
cha,
aip,
10pt,
]{revtex4-1}

\usepackage[utf8]{inputenc}
\usepackage{graphicx}
\usepackage{braket}
\usepackage{amssymb}
\usepackage{amsmath} 
\usepackage{siunitx}

\pdfoutput=1

\begin{document}

\title{Vacuum-Induced Saturation in Plasmonic Nanoparticles}

\author{Felix Stete}
\affiliation{Institut für Physik \& Astronomie, Universität Potsdam, Karl-Liebknecht-Str. 24-25, 14476 Potsdam, Germany}
\affiliation{School of Analytical Sciences Adlershof (SALSA), Humboldt-Universität zu Berlin, Unter den Linden 6, 10999 Berlin, Germany}
\author{Wouter Koopman}
\affiliation{Institut für Physik \& Astronomie, Universität Potsdam, Karl-Liebknecht-Str. 24-25, 14476 Potsdam, Germany}
\author{Carsten Henkel}
\affiliation{Institut für Physik \& Astronomie, Universität Potsdam, Karl-Liebknecht-Str. 24-25, 14476 Potsdam, Germany}
\author{Oliver Benson}
\affiliation{Institut für Physik, Humboldt-Universität zu Berlin, Newtonstraße 15, 12489 Berlin, Germany}
\author{Günter Kewes}
\affiliation{Institut für Physik, Humboldt-Universität zu Berlin, Newtonstraße 15, 12489 Berlin, Germany}
\author{Matias Bargheer}
\affiliation{Institut für Physik \& Astronomie, Universität Potsdam, Karl-Liebknecht-Str. 24-25, 14476 Potsdam, Germany}
\affiliation{Helmholtz Zentrum Berlin, Albert-Einstein-Str. 15, 12489 Berlin, Germany}

\begin{abstract}
Vacuum fluctuations are a fundamental feature of quantized fields. It is usually assumed that observations connected to vacuum fluctuations require a system well isolated from other influences. In this work, we demonstrate that effects of the quantum vacuum can already occur in simple colloidal nano-assemblies prepared by wet chemistry. We claim that the electromagnetic field fluctuations at the zero-point level saturate the absorption of dye molecules self-assembled at the surface of plasmonic nano-resonators. For this effect to occur, reaching the strong coupling regime between the plasmons and excitons is not required. This intriguing effect of vacuum-induced saturation (VISA) is discussed within a simple quantum optics picture and demonstrated by comparing the optical spectra of hybrid gold-core dye-shell nanorods to electromagnetic simulations.
\end{abstract}

\maketitle

\vspace{1cm}

\noindent The characteristic feature of plasmonic structures is their intense near-field, which localizes electromagnetic (EM) radiation to nanoscale volumes, far below the diffraction limit. The concentrated energy provided by these plasmonic near-fields is at the heart of groundbreaking inventions like surface enhanced Raman spectroscopy, \cite{Kneipp.1997} nanoscale lasing, \cite{Oulton.2009, Ma.2011} or plasmon-driven photosynthesis. \cite{Mukherjee.2013, Hartland.2017, Sarhan.2019} Moreover, plasmon cavities could become future building blocks for affordable quantum technologies, \cite{OBrien.2009} as placing an absorber into an extremely confined plasmon mode forms a cost-effective implementation of a room-temperature quantum-cavity system. \cite{Baranov.2018,Gross.2018} Compared to traditional ‘far-field’ cavities, the high mode concentration in the plasmonic near-fields cause extremely large light--matter coupling strengths. These permit to observe the quantum nature of the cavity--absorber coupling already at ambient conditions and despite the poor quality for typical plasmonic structures (bad cavity limit). \cite{OBrien.2009} One of the most surprising results of quantum electrodynamics (QED) is the existence of a finite-energy vacuum state in an optical cavity.\cite{Haroche.2006} Its energy may be understood from the generation and annihilation of virtual photons, which correspond to a fluctuation in the EM vacuum field. A series of prominent effects based on these fluctuations can be observed when placing an absorber into the cavity: the Lamb shift, \cite{Lamb.1947, Haroche.2006} the Purcell enhancement of spontaneous emission, \cite{Purcell.1946} and the vacuum Rabi splitting of coherent light matter coupling. \cite{Thompson.1992} All of these have already been more or less extensively studied also in the context of plasmonic nano-resonators. \cite{Zubairy.2008,Anger.2006,Gross.2018} It is less known that vacuum fluctuations can also influence the absorbance. In particular, the increased vacuum fluctuations in a cavity mode can lead to an effect we termed vacuum-induced saturation (VISA). It has been first  discussed in terms of a critical photon number below unity in the context of optical bi-stability. \cite{Rempe.1991} In the early days of cavity QED, it was moreover used as the emitter-related criterion for reaching the strong coupling regime. \cite{Kimble.1998,Rempe.1994} But, due to the low cavity losses in traditional cavities, the onset of VISA marks also the onset of strong coupling. \cite{Rempe.1994} Hence, VISA has rarely been discussed as a separate phenomenon.     

In this article, we introduce the concept of VISA in the context of plasmonic nano-resonators. As plasmonic cavities have intrinsically low quality factors $Q$, VISA can occur before the onset of the strong coupling regime. Indeed, we demonstrate the presence of VISA in simple colloidal metal-core--dye-shell nanoparticles at room temperature by showing that the optical spectra of these systems are only correctly reproduced when taking VISA into account. In particular, the inclusion of VISA removes the discrepancy between experiments \cite{Lekeufack.2010, Melnikau.2016} and previously published simulations based on non-saturating models. \cite{Antosiewicz.2014, Sukharev.2011} The latter predict a pure shell mode that is characterized by an additional absorption resonance, which has never been observed experimentally. Inclusion of saturation into the model removes this rogue resonance and permits us to describe the extinction spectra with a remarkable accuracy. To further strengthen the argument for the influence of VISA, we shown an excellent fit for changing particle size with a fixed set of parameters. 

\vspace{5mm}

\begin{figure*}[t]
    \centering
    \includegraphics{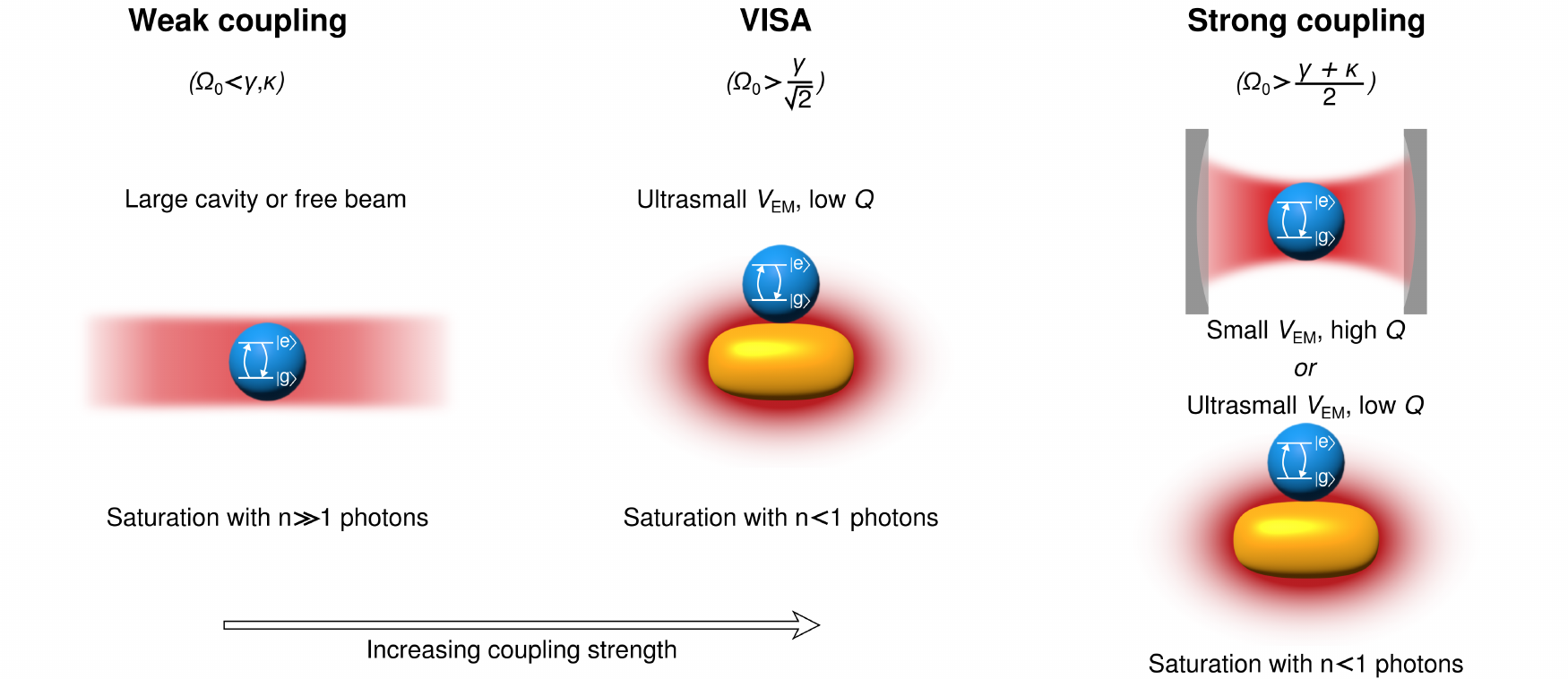}
    \caption{Sketch of different situations in which saturation occurs. Left: For low $\Omega_0$, saturation can only occur in a many photon situation ($n \gg 1$). Center: If $\Omega_0$ exceeds the absorbers linewidth $\gamma$, saturation occurs with less than one photon on average ($n < 1$). Right: Saturation is also present in the strong-coupling regime, but here, $\Omega_0$ additionally exceeds the mean of $\gamma$ and the cavity linewidth $\kappa$. In high quality mirror cavities, the onset of VISA nearly coincides with the onset of the strong-coupling regime. For plasmonic cavities, the onset of VISA occurs at lower $\Omega_0$ than the onset of the strong coupling regime.}
    \label{fig:GraphicalAbstract}
\end{figure*}

\section*{Vacuum Saturation} \vspace{-3.5mm}
 \noindent In many cases, the optical properties of atoms or molecules can be sufficiently described by a quantum two-level system. In contrast to a classical oscillator, this quantized system can be saturated by a strong EM field. This nonlinear effect becomes relevant when the intensity of the EM-mode $I$ exceeds the saturation intensity $I_\textrm{sat}$ of the absorber: \cite{Novotny.2006,Shen.1984,Grynberg.2010}
\begin{equation}
   \frac{I}{I_\textrm{sat}} = 2\cdot\frac{2I}{c \epsilon_0} \frac{\mu^2}{\hbar^2} \cdot \frac{1}{\gamma^2} = 2 \cdot \frac{ \Omega^2}{\gamma^2}\:.
    \label{eq:SaturationField}
\end{equation}
Here, $\mu$ is the transition dipole moment. From this formula, it is obvious that saturation occurs, if the light-matter coupling energy, expressed in terms of the Rabi frequency $\Omega = \mu \cdot E / \hbar $, exceeds the bare absorbers' linewidth $\gamma$. The necessary strong light-matter coupling is conventionally achieved by placing the two-level system in an intense laser beam (Figure \ref{fig:GraphicalAbstract}, left). Plasmonic nanoparticles can provide similar light intensities by concentrating EM modes to ultra-small volumes. Interestingly, if the particle sizes are chosen small enough, the intensity necessary to achieve saturation corresponds to an average of less than one photon in the cavity (Figure \ref{fig:GraphicalAbstract}, center). 

But how can saturation occur with less than one photon? $\Omega$ depends on the field strength $E$. In the vacuum state $\ket{0}$ of a cavity mode, which is defined as absence of photons in this mode, the expectation value of the electric field vanishes, as $\left< E \right> = \bra{0} E \ket{0} = 0$. However, saturation is an intensity related phenomenon. And contrary to its field strength, the intensity of the ground-state fluctuations does not vanish on average as \cite{Grynberg.2010}
\begin{equation}
   I_\textrm{vac}  = \frac{1}{2} c \epsilon_0 \bra{0} E^2 \ket{0}  =  \frac{\hbar \omega_\textrm{EM} c}{4 V_\textrm{EM}} \:.
   \label{eq:VacuumField}
\end{equation}
Here, $V_\textrm{EM}$ represents the cavity's EM mode volume and $\omega_\textrm{EM}$ is the mode frequency. The intensity of the vacuum fluctuations is thus largely determined by the geometrical confinement of the EM mode, as quantified by $V_\textrm{EM}$. As for plasmonic cavities $V_\textrm{EM}$ is extremely small, the vacuum intensity is accordingly high. Inserting the vacuum intensity from eq. \ref{eq:VacuumField} into eq. \ref{eq:SaturationField} gives:
\begin{equation}
    \frac{I_\textrm{vac}}{I_\textrm{sat}} = 2\cdot\frac{\hbar\omega_\textrm{EM}}{2V_\textrm{EM}}\frac{\mu^2}{\epsilon_0}\frac{1}{\gamma^2}=
    2\frac{\Omega_0^2}{\gamma^2}\:.
    \label{eq:CouplingStrength}
\end{equation}

\noindent In contrast to eq. \ref{eq:SaturationField}, the light--matter coupling is now given by the QED derived coupling energy for zero photons, the vacuum Rabi frequency $\Omega_0$. To arrive from eq. \ref{eq:CouplingStrength} to eq. \ref{eq:SaturationField}, $\Omega_0$ has to be multiplied by the number of photons: $\Omega^2=\Omega_n^2=\Omega_0^2 \cdot (n+1)$. We can thus interpret the situation of $I_\textrm{vac}>I_\textrm{sat}$ as vacuum induced saturation (VISA) of the absorber.

The criterion for the occurrence of VISA ($\Omega_0^2 > \gamma^2/2$) is closely related to the criteria for the onset of the strong light-matter coupling regime (Figure\,\ref{fig:GraphicalAbstract}, right). However, it makes no demands concerning the cavity quality. Thus, for absorbers coupled to low quality plasmonic cavities, with a cavity linewidth, $\kappa$, that by far exceeds $\gamma$, VISA can occur well outside the strong coupling regime.

\section*{Susceptibility of Saturated Absorbers}\vspace{-3.5mm}
\noindent The notion that vacuum fluctuations can saturate an optical transition seems odd, at the very least. This is because one typically conceptualizes saturation as excitation of a significant fraction of absorbers in an ensemble. This way of thinking is intuitive in many cases, on a quantum level it is however incorrect. Saturation already occurs on the single absorber level. \cite{Shen.1984} From the perspective of the absorber, the near-resonant photonic mode changes the potential energy landscape of the molecular states. \cite{Thomas.2019,Hutchison.2012} The adjustment to this new landscape modifies the absorber's polarizability (actually, it can also modify the absorbers quantum states \cite{Haroche.2006} through Stark effect and cavity Lamb shift). Consequently, a low-intensity beam probing the absorber will perceive a decreased absorber-susceptibility $\chi$. For an absorber placed in a time harmonic cavity mode, $E=E_0 \cdot \cos{(\omega t)}$, $\chi$ as perceived by a probe field is:\cite{Grynberg.2010,Torma.2015} 
\begin{equation}
    \chi(\omega) = \frac{f\omega_0}{2} \frac{\omega_0-\omega+i\frac{\gamma}{2}}{(\omega_0-\omega)^2+\frac{\gamma^2}{4}+\frac{\Omega^2}{2}}\:.
    \label{eq:chi_twolevel}
\end{equation}
where $\omega_0$ describes the absorber's resonance frequency and $f$ the dimensionless oscillator strength. This expression is very similar to the classical linear Lorentz-oscillator resonance, \cite{Grynberg.2010, Torma.2015} frequently used to model resonances in dielectric materials. \cite{Quinten.2011} The saturation non-linearity that distinguishes eq. \ref{eq:chi_twolevel} from its linear counterpart, originates from the appearance of the light-matter coupling energy, in form of the Rabi frequency $\Omega^2/2$, in the denominator of eq. \ref{eq:chi_twolevel}. \cite{Torma.2015} 

Traditionally, the high field intensities that are required for saturation are generated by intense laser beams.\cite{Shen.1984} Since in this "many-photon" case, $\Omega$ is mostly calculated via the electric field strength, the saturation seemingly vanishes for vanishing fields. However, for low photon numbers the quantum “graininess” of the mode becomes apparent and we must use the QED derived coupling parameter $\Omega=\sqrt{n+1} \cdot \Omega_0$. As this $\Omega$ has the finite ground-state value $\Omega_0$, \cite{Torma.2015} the saturation does not vanish, either. Typically, for $n<1$ the saturation is low enough to be safely ignored. In plasmonic cavities, on the other hand, $I_\textrm{vac}$, and thus also $\Omega_0$, is sufficiently high to cause a measurable saturation effect: VISA. In other words, the polarizability of the absorber is modified by the intensity fluctuations in the ground-state cavity mode. It is important to keep in mind, that VISA only modifies the susceptibility of an absorber. The transition to the excited state of the absorber is only realized, if the absorber is probed by a “real” photon, e.g. in a UV/Vis spectrometer. The measured absorption spectrum in this case can be interpreted as interaction of the probe photon with the vacuum fluctuations of the cavity on the absorber. VISA is by far not the only example in which this mixing of vacuum fluctuations with an excited quantum system leads to an observable physical effect. Some prominent examples for this behavior are parametric fluorescence, spontaneous emission, or vacuum Rabi splitting, which are the “vacuum versions” of parametric amplification, stimulated emission, and strong-field Rabi splitting. \cite{Grynberg.2010,Shen.1984}

\section*{Saturation in Hybrid Nanorods.}\vspace{-3.5mm}
\noindent The experimental investigation of VISA is generally challenging. To detect the characteristic spectral broadening and intensity reduction, the undisturbed absorbers' resonance must be precisely known. However, VISA can also have more qualitative effects. For example, VISA may have prevented the coupling of extremely narrow-linewidth emitters to plasmonic resonators,\cite{Chen.2013} as narrower exciton linewidth causes a significant saturation. Another example are EM simulations of hybrid metal-core--dye-shell nanoparticles. These predict a pure absorber resonance\cite{Antosiewicz.2014} that is however not observed experimentally. In the following, we will show that including VISA in EM simulations suppresses this rogue resonance, reconciling the simulations with the experimental observations.  

\begin{figure}[t]
    \centering
    \includegraphics{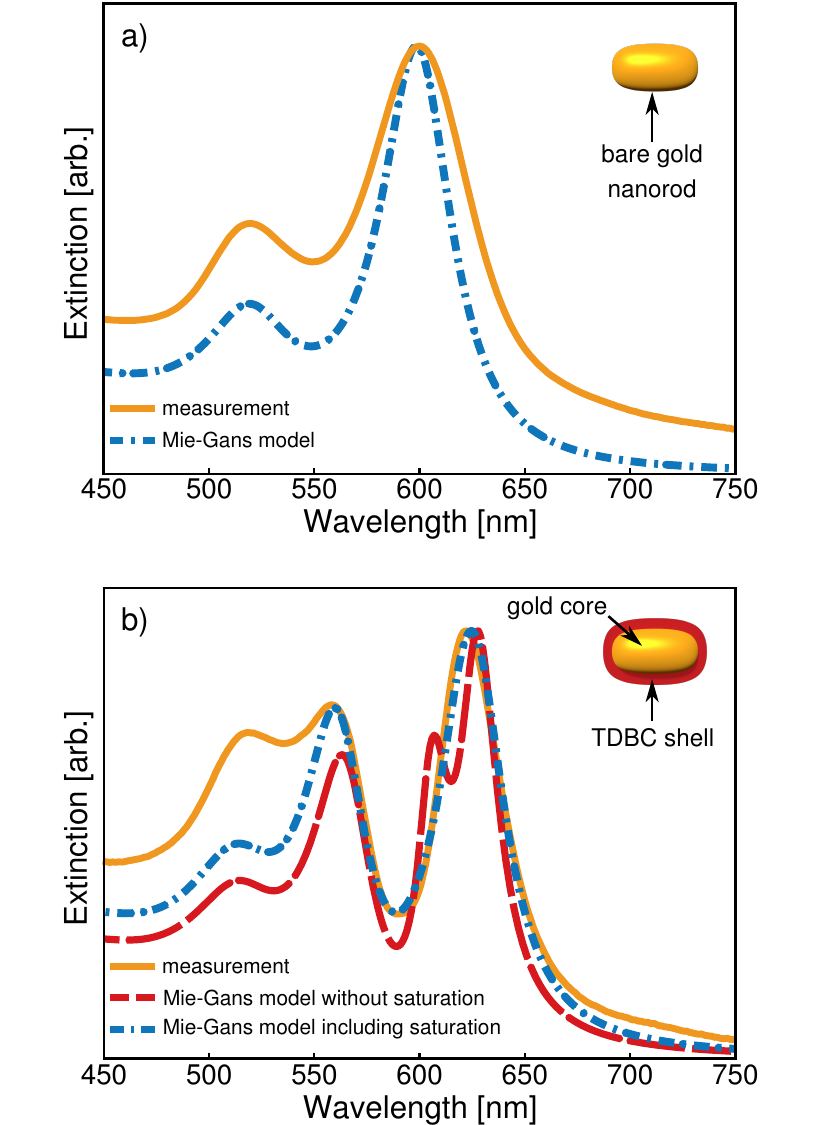}
    \caption{(a) Extinction spectrum of bare gold nanorods as measured (solid orange line) and simulated with Mie--Gans theory (dash-dotted blue line). (b) Extinction spectrum of gold nanorods coated with TDBC as measured (solid orange line), simulated with a classical shell susceptibility (dashed red line) and a shell susceptibility including VISA (dash-dotted blue line).}
   \label{fig:AuNR25-600}
\end{figure}

Mie--Gans theory is a well-known method for calculating the response of a layered nanorod system to EM excitation. \cite{Bohren.2008}
It allows the derivation of the system's extinction from the particle's dimensions (long and short axis diameters and shell thickness) and the permittivities $\epsilon$ of the involved materials.
Figure\,\ref{fig:AuNR25-600}a presents the Mie--Gans simulated extinction spectrum of a gold nanorod (dash-dotted blue line) in comparison to an experimentally measured spectrum of an aqueous nanorod dispersion (orange line). 
The short axis diameter of the particles was determined by transmission electron microscopy (TEM) to be 18\,nm, the long axis is left adjustable to fit the plasmon resonance. A 1\,nm thick citrate capping layer is included \cite{Park.2014} (details on the model can be found in the supporting information).
The two resonances in both simulation and experiment correspond to the transverse (520\,nm -- short axis) and the longitudinal plasmon mode (600\,nm -- long axis) of the rod. The discrepancy in the magnitude of the transverse resonance is a known issue in EM nanorod simulations. \cite{Myroshnychenko.2008, Huanjun.2013} We mainly attribute the differences to agglomerations of rods and secondary particles like spheres in the ensemble spectrum.

The experimental spectrum changes considerably after the gold particles are coated with a layer of the J-aggregate forming dye TDBC (orange line in Figure\,\ref{fig:AuNR25-600}b). In particular, the coupling between core and shell induces the longitudinal plasmon resonance to split into two new resonances. 

In the simulation, the shell permittivity $\epsilon_\textrm{shell}$ is now described by that of the dye material. In a classical simulation that does not consider VISA, this means $\epsilon_\textrm{shell} = \epsilon_\infty + \chi_\textrm{class}$ with $\chi_\textrm{class}$ corresponding to the classical Lorentz susceptibility and $\epsilon_\infty$ introducing the correction for off-resonant absorber transitions of higher energy. \cite{Antosiewicz.2014}

Such a classical simulation can reproduce the splitting of the longitudinal mode with good accuracy (dashed red line in Figure\,\ref{fig:AuNR25-600}b). However, the simulation shows a striking difference compared to the measurement: an additional third peak emerges close to the uncoupled absorber resonance frequency. This resonance has been described in several theoretical works \cite{Antosiewicz.2014, Sukharev.2011} on metal-core--dye-shell nanoparticles. It corresponds to an absorption in the shell without excitation of the metal core. To our knowledge, it was however never observed experimentally.
    
For the simulation in Figure\,\ref{fig:AuNR25-600}b, we chose the emitter parameters in accordance with literature. We used $\epsilon_\infty = 1.7$, \cite{Antosiewicz.2014} an absorber linewidth of $\hbar \gamma = 47\,\textrm{meV}$, \cite{Stete.2018} as well as an absorber resonance of $\hbar \omega_0= 2.0\,\textrm{eV}$ ($612\,\textrm{nm}$) and an oscillator strength of $f=0.08$. \cite{Zengin.2015} The shell thickness of $3\,\textrm{nm}$ was determined by TEM. Altering these parameters, also well outside the range found in literature, did not remove this peak while keeping the experimentally observed mode separation.

The presented model can be reconciled with the experimental data by including VISA in the shell permittivity (blue dash-dotted line in Figure\,\ref{fig:AuNR25-600}b). The prerequisite for a saturable quantum system is that the excitation to the first excited state differs energetically from the excitation of the second excited state. \cite{Shen.1984} Because of the high coupling strengths between the absorbers and the cavity, the ensemble of absorbers acts coherently like a giant oscillator. \cite{Haroche.2006} This situation is that of a single J-aggregate, \cite{Kobayashi.2012} for which the absorption of the two-exciton state is indeed energetically shifted compared to the one exciton absorption. As J-aggregate ground-state bleaching was observed in ensembles, \cite{Chakrabarti.1998,Spano.1995} as well as in core--shell geometries,\cite{Fofang.2011} we approximate the shell as saturable two-level system. 

We replace the clasical susceptibility in $\epsilon_\textrm{shell}$ with $\chi(\omega)$ of eq.\,\ref{eq:chi_twolevel} containing $\Omega_0$. While an \textit{ab initio} calculation of $\Omega_0$ is in principle possible within the framework of quasi normal modes, \cite{Lalanne.2018, Kewes.2018} it is very sensitive to the exact material parameters and the geometry. To prevent ambiguities, we use instead experimental values for $\Omega_0$, measured by variation of the dielectric environment.\cite{Stete.2017}  

Using $\hbar \Omega_0 = 85\,\textrm{meV}$, the simulated extinction spectrum including VISA reproduces the experimental spectrum almost perfectly (dash-dotted blue line in Figure\,\ref{fig:AuNR25-600}b), except for the previously mentioned issue with the magnitude of the transverse resonance. We used the same set of parameters as for the unsaturated susceptibility, except for a slightly higher oscillator strength of $f=0.11$. \cite{Zengin.2015} The striking agreement between simulation and experiment in Figure\,\ref{fig:AuNR25-600}b presents strong evidence for the presence of VISA.

Note that the experimental spectrum was recorded in a conventional transmission spectrometer in the low excitation limit.
That means, the average time between two consecutive photons reaching a particle exceeded by far the average lifetime of the plasmon and a lowering of probe light intensity had no influence on the spectrum.
In other words, a single photon in the spectrometer can already probe the saturation, which therefore must be a result of a ground-state modification. This situation is similar to vacuum Rabi splitting, where one photon is sufficient to see the mode splitting.

\begin{figure*}
    \centering
    \includegraphics{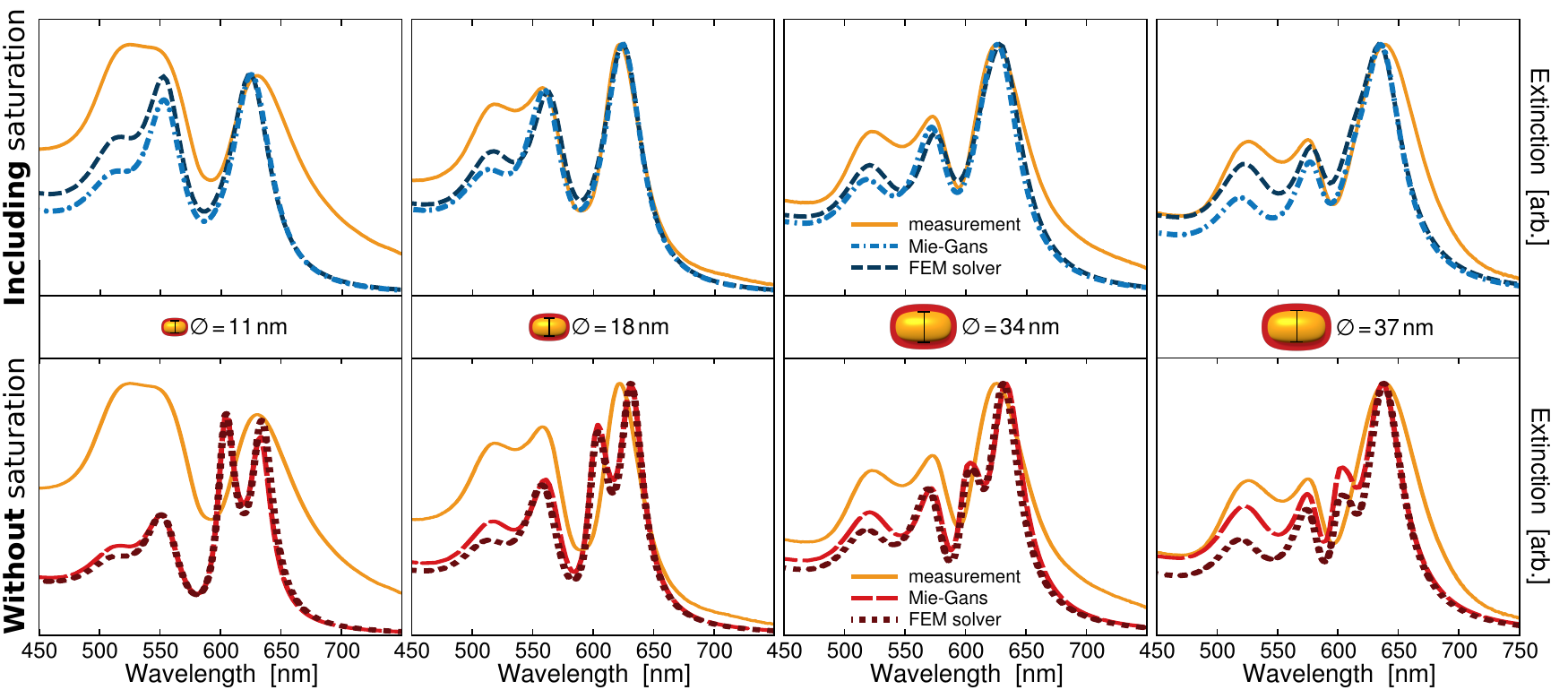}
    \caption{The upper row shows the extinction spectra of TDBC-coated gold nanorods of different sizes as measured (orange solid lines) and simulated with to the Mie--Gans model (dash-dotted blue lines) and FEM simuations (dashed dark blue lines). Both models include VISA. All particles have the same aspect ratio but a varying transverse diameter of $11\,\textrm{nm}$, $18\,\textrm{nm}$, $34\,\textrm{nm}$, and $37\,\textrm{nm}$ (from left to right). The simulations included VISA. The lower row presents the same measured spectra for the same particles (orange lines) with Mie--Gans-simulated (dashed red lines) and FEM--simulated (dotted dark red lines) that did not include saturation.}
    \label{fig:TDBC-AuNR_all}
\end{figure*}

We further tested the VISA model by simulating particles with different particle sizes corresponding to different coupling energies. \cite{Stete.2018} In addition to the analytical Mie--Gans calculation, we now also used a numerical FEM solver to determine whether the exact particle shape influences the outcome. In both calculations we used the particle dimensions as determined by TEM. Moreover, we used the same parameters set for TDBC mentioned earlier, except for $\Omega_0$, which we varied according to the measured mode splittings.  

The upper row in Figure\,\ref{fig:TDBC-AuNR_all} presents experimental and calculated spectra for particles with a transverse diameter of $11\,\textrm{nm}$, $18\,\textrm{nm}$, $34\,\textrm{nm}$, and $37\,\textrm{nm}$, which support a coupling strength $\Omega_0$ of $100\,\textrm{meV}$, $85\,\textrm{meV}$, $77\,\textrm{meV}$, and $65\,\textrm{meV}$. The agreement between measured spectra and model is remarkable. Mie--Gans and FEM simulations are consistent, no shell resonances appear at the positions of the "bare" absorber and in all cases, the split resonance is very well reproduced. The slight mismatch for the smallest particles is probably caused by the increased surface damping in such small particles, which modifies the permittivity of gold. \cite{Stoller.2006} 

The lower row in Figure\,\ref{fig:TDBC-AuNR_all} presents Mie--Gans and FEM simulations for the same particles as in the upper row, using the same parameter set, except for using $\chi_\textrm{class}$ in in the shell permittivity. In all cases, the additional absorber peak emerges. As the saturation increases with coupling strength, the rough resonance becomes more pronounced for higher coupling strength. However, we emphasize that due to the large plasmon bandwidth, only the smallest particles support strong coupling. \cite{Rempe.1994, Khitrova.2006} This confirms that only the relation between $\Omega_0$ and $\gamma$ is the relevant criterion and VISA can indeed occur in the bad cavity limit.

\section*{Conclusion}\vspace{-3.5mm}
\noindent In conclusion, we introduced the concept of vacuum induced saturation (VISA) of absorbers in plasmonic nanocavities. VISA is caused by the intensity fluctuations of an electromagnetic mode, which are inversely proportional to its mode volume. Plasmonic modes with sub-diffraction volumes hence possess extreme vacuum intensity fluctuations. These cause a reduction in the polarizability of the absorber, which effectively saturates its resonance. We presented evidence for VISA by the example of simple gold-core--dye-shell nanorods.

We make our predictions accessible to numerical modeling by including VISA into the susceptibility, which is the essential parameter in the context of Mie scattering. In our experimental system of core--shell nanorods with a plasmonic core and an excitonic absorber shell, classical models predict a resonant mode of the particle shell that has never been  observed in experiments. This mode is indeed suppressed when VISA is included in the model. We take the fact that the inconsistencies of experiment and theory can be removed elegantly with a single parameter, the coupling strength, as a strong evidence that VISA occurs in plasmon absorber systems and can be incorporated into the modeling as specified. The evidence is further supported by the fact that only the coupling strength has to be adjusted for simulating rods of different size.

The presented insights suggest that QED effects can be relevant in plasmonic systems well outside the strong coupling regime and at least a part of the rich physics of quantum non-linearties, \cite{Wang.2019} such as large single photon phase shifts, strongly non-classical photon statistics of scattered light or applications such as single photon logical gates,\cite{Turchette.1995} might be accessible with less effort in simple and cost-effective systems. Moreover, the extremely short lifetimes of plasmonic excitations can be a potential benefit for ultrafast photon routing or switching in high-rate quantum information processing.

\section*{Supporting Information}\vspace{-3.5mm}
\noindent Details on sample preparation and characterization, measurements of $\Omega_0$, analytical and numerical simulations are presented in the supporting information.

\section*{Acknowledgement}\vspace{-3.5mm}
\noindent We thank Ferenc Liebig and Zdravko Kochovski for recording the TEM images for particle size and shell thickness determination. Additionally, G.K., O.B. and C.H. acknowledge financial support by the Deutsche Forschungsgemeinschaft (DFG, German Research Foundation) via project 182087777 - SFB 951 (G.K. and O.H.) and projects Schm-1049/7-1 and Fo 703/2-1 (both C.H.). Also, F.S. wants to thank the DFG for financial support via the graduate school SALSA. Moreover, we would like to thank the JCMwave team for fruitful discussions.

\section*{Competing Interests}\vspace{-3.5mm}
\noindent The authors declare no competing interest.

\bibliography{references.bib}

\end{document}


\title{Supporting Information for "Vacuum-Induced Saturation in Plasmonic Nanoparticles"}

\author{Felix Stete}
\affiliation{Institut für Physik \& Astronomie, Universität Potsdam, Karl-Liebknecht-Str. 24-25, 14476 Potsdam, Germany}
\affiliation{School of Analytical Sciences Adlershof (SALSA), Humboldt-Universität zu Berlin, Unter den Linden 6, 10999 Berlin, Germany}
\author{Wouter Koopman}
\affiliation{Institut für Physik \& Astronomie, Universität Potsdam, Karl-Liebknecht-Str. 24-25, 14476 Potsdam, Germany}
\author{Carsten Henkel}
\affiliation{Institut für Physik \& Astronomie, Universität Potsdam, Karl-Liebknecht-Str. 24-25, 14476 Potsdam, Germany}
\author{Oliver Benson}
\affiliation{Institut für Physik, Humboldt-Universität zu Berlin, Newtonstraße 15, 12489 Berlin, Germany}
\author{Günter Kewes}
\affiliation{Institut für Physik, Humboldt-Universität zu Berlin, Newtonstraße 15, 12489 Berlin, Germany}
\author{Matias Bargheer}
\affiliation{Institut für Physik \& Astronomie, Universität Potsdam, Karl-Liebknecht-Str. 24-25, 14476 Potsdam, Germany}
\affiliation{Helmholtz Zentrum Berlin, Albert-Einstein-Str. 15, 12489 Berlin, Germany}

\maketitle

\noindent Here, we give a concise summary of the relevant experimental methods followed by details on the simulations of the extinction spectra.

\section{Hybrid Gold--TDBC Nanorods} \vspace{-3.5mm}

\noindent We fabricated hybrid nanorods along the lines of a procedure first published by Lekeufack et al. \cite{Lekeufack.2010} In short, 5,5',6,6'-tetrachloro-1,1'-diethyl-3,3'-di(4sulfobutyl)ben\-zi\-midazolo\-carbo\-cyanine (TDBC, from FEW Chemicals) was dissolved in NaOH ($c_{\textrm{NaOH}}=\SI{e-5}{mol/l}$) to a concentration of $c_{\textrm{TDBC}} \approx \SI{1}{mmol/l}$. Gold nanorods with a citrate ligand
were added to the TDBC solution with a ratio of 1:1. After $\SI{48}{h}$, the TDBC coated nanorods were washed by centrifugation.

In the experiments, we used rods of four different sizes, with short axis diameters of nominally $10\,\textrm{nm}$, $25\,\textrm{nm}$, $40\,\textrm{nm}$, and $50\,\textrm{nm}$. The length of the long axes was chosen such that the aspect ratios was approximately 2 for all rods and hence, the resonance of the longitudinal plasmon mode was around 600\,nm, which is close to the absorbers' resonance.

\section{Characterization}\vspace{-3.5mm}

\begin{figure}[h]
    \centering
    \includegraphics[width=0.65\textwidth]{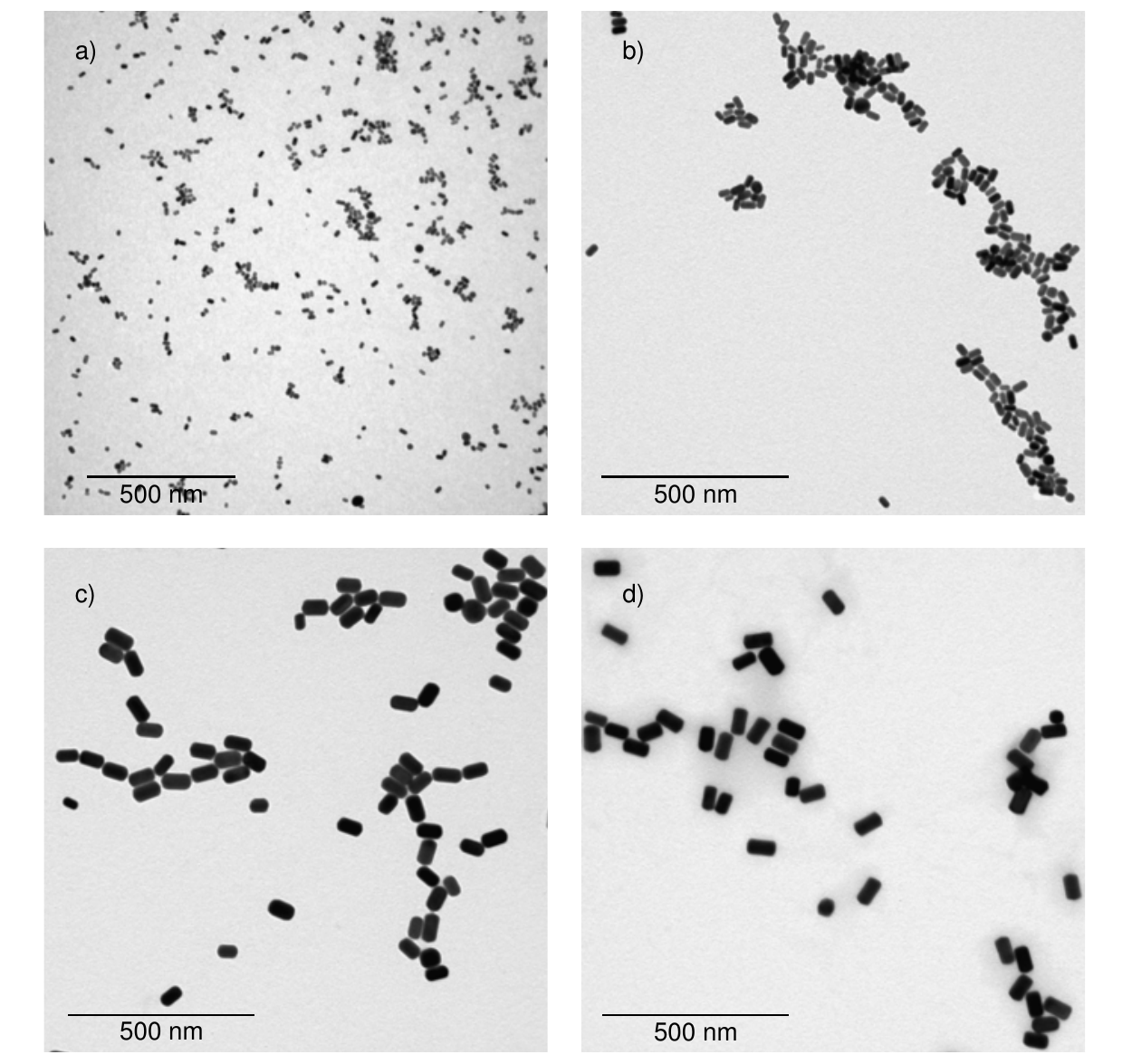}
    \caption{Typical TEM images of the gold nanorods. The transverse diameters were retrieved to be (a) $11\,\textrm{nm}$, (b) $18\,\textrm{nm}$, (c) $34\,\textrm{nm}$, and (d) $37\,\textrm{nm}$}
   \label{fig:TEMOverview}
\end{figure}

\noindent For higher precision of the simulations, we measured the exact particle dimensions and the coating layer thickness by transmission electron microscopy (TEM).
In a JEM-1011 (JEOL Japan), we recorded ensemble images for each particle size and retrieved the respective long and short axis diameters.
Representative images are presented in Figure\,\ref{fig:TEMOverview}. The results of the size measurements are summarized in Table\,\ref{tab:ParticleSizes}.

\vspace{5mm}
\begin{table}
\centering
\begin{tabular}{ccccc}
  \multicolumn{1}{l}{Nominal short}      &\hspace{2mm} & \multicolumn{1}{l}{Recorded short} &\hspace{2mm}  & \multicolumn{1}{l}{Recorded long}    \\
  \multicolumn{1}{l}{axis length [nm]}  & & \multicolumn{1}{l}{axis length [nm]} && \multicolumn{1}{l}{axis length [nm]}  \vspace{3mm} \\
  10   & & 11 $\pm$ 1  && 23 $\pm$ 4 \\
  25   & & 18 $\pm$ 3  && 35 $\pm$ 6 \\
  40   & & 34 $\pm$ 4  && 64 $\pm$ 7 \\
  50   & & 37 $\pm$ 3  && 71 $\pm$ 5 \\
\end{tabular}
\caption{Summary of measured particle sizes from TEM images.}
\label{tab:ParticleSizes}
\vspace{5mm}
\end{table}

\begin{figure}[t]
    \centering
    \includegraphics[width=0.4\textwidth]{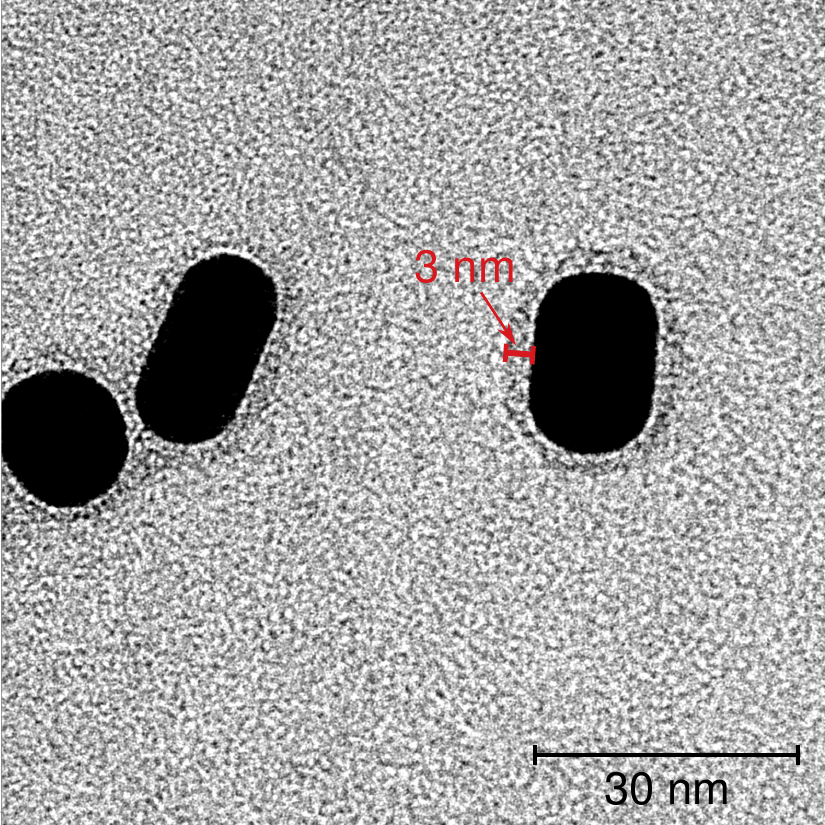}
    \caption{Typical TEM image with of the TDBC coated gold nanorods, with the 3\,nm TDBC monolayer clearly visible. The image shows the presence of spherical particles, which influence the high energy plasmon spectrum.}
   \label{fig:TEM}
\end{figure}

\noindent In a separate measurement, we employed another transmission electron microscope (JEM-2100, JEOL Germany) to determine the thickness of the TDBC coating. A typical image of a TDBC-coated gold nanorod is presented in Figure\,\ref{fig:TEM}. We retreived a shell thickness of 3\,nm. This value agrees well with previously published values for TDBC monolayers. \cite{Zengin.2015}

\section{Measurement of Vacuum Rabi Frequency} \vspace{-3.5mm}

\noindent We measured the plasmon--exciton coupling strength using a dielectric shift technique. For that purpose, we deposited gold nanorods on a substrate and subsequently covered the samples with thin layers of polyelectrolytes. \cite{Stete.2017} This provided a way to tune the particle plasmon resonance, and the anticrossing with the absorber resonance directly yielded $\Omega_0$ for the particles investigated here. \cite{Stete.2018}

\noindent Glass substrates were modified by spin-coating a layer of polyelectrolyte layers poly-ethy-leneimine (PEI) and a layer of  poly-sodium 4-styrenesulfonate (PSS). The TDBC coated nanorods were drop-cast onto the substrates and given $\SI{12}{h}$ time for adsorption before residuals were washed away with purified water. A series of samples with different thicknesses of  spin coated multilayers of PSS and poly-allylamine hydrochloride (PAH) on top of the particles was created. All mentioned chemicals were purchased form Sigma Aldrich. For each sample, an absorption spectrum was recorded. In order to separately assess the shift of the plasmon spectrum in the absence of TDBC,  each sample was photobleached with a continuous-wave laser working at $\SI{532}{nm}$ with a power of $\SI{10}{W}$. \cite{Stete.2017} Again, a spectrum was recorded of each sample after the bleaching process to determine the spectral position of the 'bare' plasmon resonance. The (longitudinal) maximum positions of the coupled spectra were eventually plotted against the (longitudinal) maximum positions of the bleached spectra. The resulting anticrossing could be fitted \cite{Stete.2017, Stete.2018} by
\begin{equation}
\omega_\pm = \frac{\omega_\textrm{pl} + \omega_\textrm{ex}}{2} \pm \sqrt{\frac{(\omega_\textrm{pl}-\omega_\textrm{em})^2}{4} + \Omega_0^2}
\end{equation}
with $\omega_\pm$ as upper and lower branch of the anticrossing and $\omega_\textrm{i}$ as resonance of plasmon ($i=\textrm{pl}$) or exciton ($i=\textrm{ex}$). The fit directly yields the vacuum Rabi frequency $\Omega_0$. With this procedure, we experimentally obtained $\hbar \Omega_0$ for the four particle sizes to be $116\, \textrm{meV}$, $78\, \textrm{meV}$, $73\, \textrm{meV}$, and $54\, \textrm{meV}$ (from smallest to largest particles, respectively). \cite{Stete.2018}

The slight deviation of the values of $\Omega_0$ used in the article and the measured ones presented here can easily be explained by the difference of chemical environments in the two sets of experiments (in one case, particles were immobilised on a substrate, in one case, the particles wee dispersed in solution).

\section{Analytical Simulations} \vspace{-3.5mm}
\noindent The extinction spectra of the hybrid gold-core--dye-shell nanorods were calculated according to the established Mie--Gans procedure as detailed by Bohren and Huffmann. \cite{Bohren.2008}

Mie--Gans theory provides an expression for the dipolar polarizability $\alpha_{0,j}$ of a core--shell nanospheroid along the long ($j=a$) or short ($j=b$) axis. With the permittivities of the core ($\epsilon_\textrm{core}$), the shell ($\epsilon_\textrm{shell}$) and the surrounding medium ($\epsilon_\textrm{med}$), it reads as \cite{Bohren.2008, Zengin.2013}
\begin{equation}
\begin{split}
\alpha_{0,j} =  V
\frac{(\epsilon_\textrm{\textrm{shell}}-\epsilon_\textrm{\textrm{med}}) \epsilon_\textrm{u} + g \epsilon_\textrm{\textrm{shell}}(\epsilon_\textrm{\textrm{core}}-\epsilon_\textrm{\textrm{shell}})}
{(\epsilon_\textrm{\textrm{med}}+L^{(2)}_j (\epsilon_\textrm{\textrm{shell}}-\epsilon_\textrm{\textrm{med}}))\epsilon_\textrm{u} + g L^{(2)}_j \epsilon_\textrm{\textrm{shell}}(\epsilon_\textrm{\textrm{core}}-\epsilon_\textrm{\textrm{shell}})}\:.
\end{split}
\label{eq:Polarisability}
\end{equation}
Here, $\epsilon_\textrm{u} = \epsilon_\textrm{\textrm{shell}} + (\epsilon_\textrm{\textrm{core}} - \epsilon_\textrm{\textrm{med}})(L^{(1)}_j - g L^{(2)}_j)$ with the depolarization factors $L^{(1,2)}_j$ for the inner (1) and outer (2) spheroid and $g$ as the inner spheroid's volume fraction of the total volume $V$. 

We account for the finite particle size by applying the modified wavelength approximation \cite{Kelly.2003, Zengin.2013} to obtain a modified polarizability $\alpha$ as
\begin{equation}
\alpha_j = \alpha_{0,j} \left( 1 - i \frac{\alpha_{0,j}}{6 \pi \epsilon_0} k^3 - \frac{\alpha_{0,j}}{4 \pi \epsilon_0} \frac{k^2}{s_j}
\right)^{-1}
\label{eq:MLWA}
\end{equation}
where $k$ is the wave vector in the medium and $s_j$ the respective semiaxis.

The extinction cross section $\sigma_\textrm{ext,j}$ can be computed straight-forwardly from this polarizability as 
\begin{equation}
\sigma_\textrm{ext,j} = \frac{k^4}{6 \pi \epsilon_0^2} |\alpha_j|^2 + \frac{k}{\epsilon_0} \text{Im}(\alpha_j)\:.
\label{eq:SigmaExt}
\end{equation}
As the particles were dissolved in solution and consequently randomly oriented, we included all three axes with a weighting factor of 1/3. \cite{Bohren.2008}

\begin{figure}[t]
    \centering
    \includegraphics[width=\textwidth]{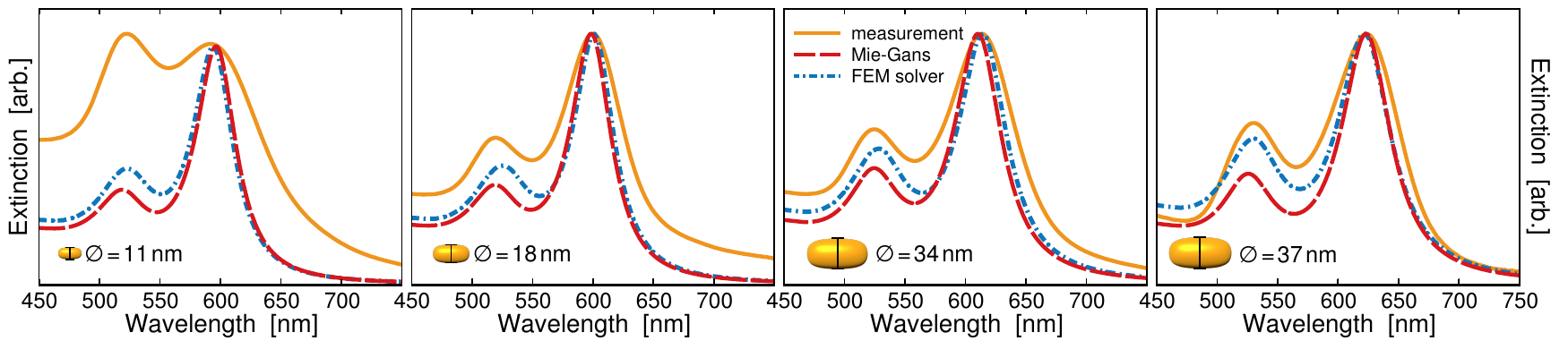}
    \caption{Measurements and simulations of the bare gold nanorods. The solid orange lines repesent the measurement data, the dashed red lines represent the analytical data retrieved from Mie--Gans theory, the dash-dotted blue lines represent the numerical simulations.}
   \label{fig:BareParticleSpectra}
\end{figure}

In the simulations, the permittivity of gold was taken from literature, \cite{Olmon.2012} while a value of $\epsilon_{\textrm{shell}} = 1.54^2$ was assumed for the 1\,nm citrate shell. \cite{Park.2014} The short axis diameter of the particles was taken from the TEM measurements. As real nanorods do not have the exact spheroidal shape assumed in the Mie--Gans model, we adjusted the long axis length such that the calculation fitted the experimental plasmon resonance. A comparison between simulations and measurements is given in Figure\,\ref{fig:BareParticleSpectra}.

\section{Numerical Simulations} \vspace{-3.5mm}
\noindent We used the commercial finite element method (FEM) solver for Maxwell’s equations JCMSuite and applied cylindrical coordinates. Instead of spheroids, we used cylinders with rounded edges to model the rods. Rod dimensions and curvatures were retrieved from TEM images. The other parameters (shell thickness, permittivities) were identical to those used in the analytical simulations as described in the text. Cross sections were calculated as the mean from three plane wave light sources, one propagating parallel and two orthogonal to the long axis of the rod, respectively. The latter two are either polarized along the short or the long axis of the rod. This average was tested to yield the same results as a significantly more time consuming average over all possible angles of incidence that must be used in principle for non-spherical particles with larger aspect ratios. The results for the bare particles are presented in Figure\,\ref{fig:BareParticleSpectra}.

\section{Classical Oscillator and Saturation Profile} \vspace{-3.5mm}

\noindent Commonly, an  absorber is approximated as a Lorentz oscillator with resonance frequency $\omega_0$ and linewidth $\gamma$. For an ensemble of oscillators, a macroscopic material model can be derived, represented by a susceptibility $\chi_{\textrm{class}}$ as \cite{Torma.2015}
\begin{equation}
    \chi_{\textrm{class}}(\omega) = \frac{f\omega_0^2}{\omega_0^2-\omega^2+i\gamma\omega} \approx \frac{f\omega_0}{2} \frac{\omega_0-\omega+i\frac{\gamma}{2}}{(\omega_0-\omega)^2+\frac{\gamma^2}{4}}
    \label{eq:chi_Lorentz}
\end{equation}
where the dimensionless oscillator strength $f$ is proportional to the absorber density. The approximation is valid for $\omega \gg \gamma$, close to resonance.

Within this linear response model, no saturation can occur. Non-linearities are included if the absorbers are treated as two-level systems in an oscillating classical electric field $E=E_0 \cos{\omega t}$. In this case, the susceptibility can be derived from the optical Bloch equations and reads \cite{Grynberg.2010,Torma.2015}
\begin{equation}
    \chi(\omega) = \frac{f\omega_0}{2} \frac{\omega_0-\omega+i\frac{\gamma}{2}}{(\omega_0-\omega)^2+\frac{\gamma^2}{4}+\frac{\Omega^2}{2}}\:.
    \label{eq:chi_twolevel}
\end{equation}
where $\Omega = \mu E_0 /\hbar$ is the Rabi frequency. It is instructive to write $\chi$ in terms of a saturation parameter $s$:
\begin{equation}\label{sat}
    \chi(\omega) = \frac{\chi_{\textrm{class}}(\omega)}{1+s(\omega)}
\end{equation}
with
\begin{equation}
    s(\omega) = \frac{\frac{\Omega^2}{2}}{(\omega_0-\omega)^2+\frac{\gamma^2}{4}}
    =
    \frac{ I }{ I_\mathrm{sat}} \frac{ \frac{\gamma^2}{4}}{ (\omega_0-\omega)^2+\frac{\gamma^2}{4} }
    \:.
\end{equation}
When $s$ is small, $\chi$ is not different from the classical solution. As soon as $s$ becomes non-negligible (around $I = I_\textrm{sat}$ near resonance), $\chi$ is reduced and for extremely high intensities pushed towards zero indicating a saturation of the absorbers.

\bibliography{references.bib}